\documentclass[12pt]{article}
\usepackage{graphicx}

\begin{document}

\author{M. L. D. Ion and D. B. Ion}
\title{{\bf PLANE\ SPDC-QUANTUM\ MIRROR}}
\maketitle

\begin{abstract}
In this paper the kinematical correlations from the {\it phase conjugated
optics ( }equivalently with {\it crossing} {\it symmetric spontaneous
parametric down conversion (SPDC) phenomena)} in the nonlinear crystals are
used for the description of a new kind of optical{\it \ }device called SPDC-%
{\it quantum mirrors. }Then,{\it \ s}ome important laws of the{\it \ plane} 
{\it SPDC-quantum mirrors} combined with usual mirrors or lens are proved
only by using geometric optics concepts. In particular, these results allow
us to obtain a new interpretation of the recent experiments on the {\it %
two-photon geometric optics}.

\smallskip

PACS: \smallskip\ 42. 50. Tv ; 42. 50. Ar ; 42. 50. Kb ; 03. 65. Bz.\ 
\end{abstract}

\section{{\bf Introduction}{\it \ }}

The {\it spontaneous parametric down conversion} (SPDC) is a nonlinear
optical process [1] in which a laser pump (p) beam incident on a nonlinear
crystal leads to the emission of a correlated pair of photons called signal
(s) and idler (i). If the {\it S-matrix crossing symmetry [2] }of the
electromagnetic interaction in the {\it spontaneous parametric down
conversion} (SPDC) crystals is taken into account, then the existence of the 
{\it direct SPDC process}

\begin{equation}  \label{1}
p\rightarrow s+i
\end{equation}
will imply the existence of the following {\it crossing symmetric processes
[3]} 
\begin{equation}  \label{2}
p+\stackrel{\_}{s}\rightarrow i
\end{equation}
\begin{equation}  \label{3}
p+\stackrel{\_}{i}\rightarrow s
\end{equation}
as real processes which can be described by the same{\it \ transition
amplitude.} Here, by $\stackrel{\_}{s}$ and $\stackrel{\_}{i}$ we denoted
the {\it time reversed} {\it photons (}or antiphotons in sense introduced in
Ref. [4]{\it )} relative to the original photons $s$ and $i$, respectively.
In fact the SPDC effects (1)-(3) can be identified as being directly
connected with the $\chi ^{(2)}$-{\it second-order nonlinear effects} called
in general {\it three wave mixing }(see{\it \ }Ref.[5]{\it ). }So,{\it \ }%
the process (1) is just the {\it inverse of second-harmonic generation, }%
while, the effects (2)-(3) can be interpreted just as emission of {\it %
optical phase conjugated replicas} in the presence of pump laser via {\it %
three wave mixing. }

In this paper a new kind of geometric optics called {\it quantum
SPDC-geometric optics} is systematically{\it \ }developed by using {\it \
kinematical correlations of the pump, signal and idler photons} from the
SPDC processes. Here we discuss only the plane quantum mirror. Other kind of
the SPDC-quantum mirrors, such as spherical SPDC quantum mirrors, parabolic
quantum mirrors, etc., will be discussed in a future paper.{\it \ }

\section{{\bf Quantum kinematical correlations}{\it \ }}

In the SPDC processes (1) the energy and momentum of photons are conserved:

\begin{equation}  \label{4}
\omega _p=\omega _s+\omega _i,\smallskip\ {\bf k_p}={\bf k_s}+{\bf k_i}
\end{equation}

Moreover, if the crossing SPDC-processes (2)-(3) are interpreted just as
emission of {\it optical phase conjugated replicas }in the presence of input
pump laser then Eqs. (4) can be identified as being the{\it \ phase matching}
{\it conditions} in the three wave mixing (see again Ref. [5]). Indeed, this
scheme exploits the second order optical nonlinearity in a crystal lacking
inversion symmetry. In such crystals, the presence of the input pump (${\bf %
E_p}$) and of the signal (${\bf E_s}${\bf )} fields{\it \ } induces in the
medium a {\it nonlinear optical polarization }(see Eqs. (26)-(27) in Pepper
and Yariv Ref.[5]) which is: $P_i^{NL}=\chi _{ijk}^{(2)}E_{pj}(\omega
_p)E_{sk}^{*}(\omega _s)\exp \{i[(\omega _p-\omega _s)t-({\bf k_p-k_s)\cdot r%
}]\}+c.c.,$ where $\chi _{ijk}^{(2)}$ is the susceptibility of rank two
tensor components of the crystal. Consequently, such polarization, acting as
a {\it source} in the {\it wave equation} will radiate a {\it new wave} $%
{\bf E_i}$ at frequency $\omega _i=$ $\omega _p-\omega _i,$ with an
amplitude proportional to ${\bf E_i^{*}}(\omega _i),$ i.e., to the{\it \
complex} {\it conjugate} of the spatial amplitude of the low-frequency probe
wave at $\omega _s.$ Then, it is easy to show that a necessary condition for
a {\it phase-coherent} cumulative buildup of {\it conjugate-field} {\it %
radiation} at $\omega _i=\omega _p-\omega _s$ is that the wave vector {\bf k$%
_i$} at this new frequency must be equal to{\bf \ k$_i$}$={\bf k}_p-{\bf k}%
_s,$ i.e., we have the phase matching conditions (4). Hence, the {\it optical%
} {\it phase conjugation by three-wave mixing } help{\it \ } us to obtain a
complete proof of the existence of the crossing reactions (2)-(3) as real
processes which take place in the nonlinear\ crystals when the {\it %
energy-momentum }(or {\it phase} {\it matching) conditions} (4) are
fulfilled.{\it \ }

Now, it is important to introduce the {\it momentum projections}, parallel
and orthogonal to the pump momentum, and to write the momentum conservation
law from (4) as follows 
\begin{equation}  \label{5}
k_p=k_s\cos \theta _{ps}+k_i\cos \theta _{pi}
\end{equation}
\begin{equation}  \label{6}
k_s\sin \theta _{ps}=k_i\sin \theta _{pi}
\end{equation}
where the angles $\theta _{pj,}$ $j=s,i,$are the angles (in crystal) between
momenta of the {\it pump} (p)$\equiv ${\it ($\omega _p,$}${\bf k}${\bf $%
_p,e_p,\mu _p)$}, {\it signal} (s)$\equiv (${\it $\omega _s,$}${\bf k}${\bf $%
_s,$}${\bf e}${\bf $_s,\mu _s)$} and {\it idler }(i)$\equiv $({\it $\omega
_i,$}${\bf k}${\bf $_i,e_i,\mu _i)$} {\it photons.} By{\bf \ e$_j$} and $\mu
_j{\bf ,\ }j\equiv p,s,i,${\bf \ }we denoted the photon polarizations and
photon helicities, respectively. Now, let $\beta _{ps},$and $\beta _{pi}$ be
the corresponding exit angles of the signal and idler photons from crystal.
Then from (6) in conjunction with Snellius law, we have 
\begin{equation}  \label{7}
\sin \beta _{ps}=n_s\sin \theta _{ps},\smallskip\ \sin \beta _{pi}=n_i\sin
\theta _{pi}
\end{equation}

\begin{equation}  \label{8}
\omega _s\sin \beta _{ps}=\omega _i\sin \beta _{pi}
\end{equation}

\section{\bf Quantum mirrors via SPDC phenomena }

(D.1) {\bf Quantum Mirror }(QM). By definition a {\it quantum mirror (QM) is
a combination of standard devices} (e.g., usual lenses, usual mirrors,
lasers, etc.) with a nonlinear crystal {\it by which one involves the use of
a variety of quantum phenomena to exactly transform ${\bf \ }$not only the
direction of propagation of a light beam but also their polarization
characteristics.}

(D.2) {\bf SPDC}-{\bf Quantum Mirror} (SPDC-QM). A {\it quantum mirrors} is
called SPDC-QM if is based on the quantum SPDC phenomena (1)-(3) in order to
transform {\it signal photons} characterized by {\it ($\omega _s,{\bf %
k_s,e_s,\mu _s)}$}${\bf \ }$into {\it idler photons} with {\it ($\omega
_p-\omega _s,{\bf k_p-}{\bf k_s,e_s^{*},-\mu _s)\equiv }$($\omega _i,{\bf %
k_i,e_i,\mu _i)}$}.

Now, since the crossing symmetric SPDC\ effects (2)-(3) can be interpreted
just as emission of {\it optical phase conjugated replicas} in the presence
of pump laser via {\it three wave mixing, }the high quality of the SPDC-QM
will be given by the following peculiar characteristics: (i){\it \ Coherence}%
:The SPDC-QM {\it preserves high coherence} between s-photons and i-photons;
(ii) {\it Distortion undoing:} The SPDC-QM {\it corrects all the aberrations}
which occur in signal or idler beam path; (iii) {\it Amplification: }A
SPDC-QM\ {\it amplifies the conjugated wave} if some conditions are
fulfilled. \ 

{\it 3.1. Plane SPDC-quantum mirrors.} The quantum mirrors can be {\it plane
quantum mirrors }(P-QM) (see Fig.1), {\it spherical quantum mirrors (S-QM),
hyperbolic quantum mirror (H-QM), parabolic quantum mirrors (PB-QM), etc.,\ }%
according with the character of incoming laser wave fronts ( {\it plane
waves,} {\it spherical waves, etc.). }Here we discuss only the{\it \ plane
SPDC-quantum mirror.} Other kind of the {\it SPDC-quantum mirrors}, such as 
{\it spherical SPDC quantum mirrors, parabolic quantum mirrors}, etc., will
be discussed in a future paper.{\it \ }

In order to avoid many complications, in the following we will work only in
the {\it thin crystal approximation}. Moreover, we do not consider here the
so called optical aberrations.

(L.1){\it \ }Law of {\it thin plane SPDC-quantum mirror}: Let BBO be a SPDC\
crystal illuminated uniform by a high quality laser pump. Let Z$_s$ and Z$_i 
$ be the distances shown in Fig.1 ( from the {\it object point} P to crystal
(point A) and from crystal (point A) to {\it image point} I$.$ Then, the
system behaves as a {\it plane mirror} but satisfying the following
important laws: 
\begin{equation}  \label{9}
\frac{Z_i}{Z_s}=\frac{\omega _i}{\omega _s}=\frac{\sin \beta _{ps} }{\sin
\beta _{pi}}=\frac{n_s\sin \theta _{ps}}{n_i\sin \theta _{pi}},\smallskip\ M=%
\frac{\omega _sZ_i}{\omega _iZ_s}=1
\end{equation}
where M is the{\it \ linear magnification }of{\it \ }the plane SPDC-quantum
mirror.

{\it 3.2. Plane SPDC-QM combined with thin lens. }The basic optical
geometric configurations of a plane SPDC-QM combined with thin lens is
presented in Figs. 2a and 2b. The system in this case behaves as in usual
geometric optics but with some modifications in the non degenerate case
introduced by the presence of the {\it plane SPDC-quantum mirror}. The
remarkable law in this case is as follows.

(L.2) {\it Law} of the{\it \ thin lens combined with a plane SPDC-QM: The
distances S ( lens-object), S'(lens-crystal-image plane), D$_{CI}$
(crystal-image plane) and f (focal distance of lens), satisfy the following
thin lens equation 
\begin{equation}  \label{10}
\frac 1S+\frac 1{S^{\prime }+(\frac{\omega _s}{\omega _i}-1)\:%
D_{CI}}=\frac 1f
\end{equation}
The SPDC-QM system in this case has the magnification M given by}

\begin{equation}  \label{11}
M=\frac{S^{\prime }+(\frac{\omega _s}{\omega _i}-1)\:D_{CI}}S=M_0+(%
\frac{\omega _s}{\omega _i}-1)\frac{D_{CI}}S
\end{equation}
In degenerate case $(\omega _s=\omega _i=\omega _p/2)$ we obtain the usual 
{\it Gauss law for thin lens} with the magnification $M_0=S^{\prime }/S$.

{\it Proof:} The proof of the predictions (10)-(11) can be obtained by using
the basic geometric optical configuration presented in Fig. 2a. Hence, the
image of the object P in the thin lens placed between the crystal and object
is located according to the Gauss law 
\begin{equation}  \label{12}
\frac 1S+\frac 1{S_1}=\frac 1f
\end{equation}
where $S_1$ is the distance from lens to image I$_1.$ Now the final image I
of the image I$_1$ in the plane SPDC-QM is located according to the law (9).
Consequently, if d is the lens-crystal distance then we have 
\begin{equation}  \label{13}
S_1=S^{^{\prime }}+(Z_s-Z_i)=S^{^{\prime }}+(\frac{\omega _s}{\omega _i}%
-1)D_{CI}
\end{equation}
since S$_1=d+Z_s,$ S'=d+Z$_i,$and D$_{CI}$ is the crystal-image distance. A
proof a the magnification factor can be obtained on the basis of geometric
optical configuration from Fig. 2b. Hence, the magnification factor is

\begin{equation}  \label{14}
M=\frac{y_I}{y_O}=\frac{y_I}{y_I^{\prime }}\cdot \frac{y_I^{\prime }}{y_O}=%
\frac{y_I^{\prime }}{y_O}
\end{equation}
since the plane SPDC-QM has the magnification $\frac{y_I}{y_I^{\prime }}=1.$
Obviously, from $\Delta PP^{\prime }V\sim \Delta I_1I_1^{\prime }V,$we get y$%
_I^{\prime }$/y$_O=S_1/S$ and then with (13) we obtain the magnification
(11).

{\it (L.3)} {\it Law of thin lens + plane SPDC-QM with the null crystal-lens
distance}

\begin{equation}  \label{15}
\frac 1S+\frac 1{\frac{\omega _s}{\omega _i}S^{\prime }}=\frac 1f \:,%
\smallskip\ M=\frac{\omega _s}{\omega _i}\frac{S^{\prime }}S
\end{equation}

{\it Proof: }Here we note that (L.4) is the particular case of (L.3) with
d=0 for which we get S$_1=Z_s,$and S'=Z$_i.$ Then from (9) and (12) we
obtain (15).

{\it 3.3. Thin lens combined with plane SPDC-QM and classical mirror.}

{\it (L.4) } {\it Law of thin lens + plane SPDC-QM +classical mirror (} see
the basic geometric optical configuration presented in Fig. 3). The
distances S ( lens-object), S$_1^{^{\prime }}$(lens-crystal-first image
plane I$_1$), S$_2^{^{\prime }}$(lens-crystal-second image plane I$_2$), D$%
_{CI_1}$ (crystal-first image plane), D$_{CI_2}$ (crystal-second image
plane) and f (focal distance of lens), must satisfy the following law{\it \ 
\begin{equation}  \label{16}
\frac 1S+\frac 1{S_1^{\prime }+(\frac{\omega _s}{\omega _i}-1) \:%
D_{CI_1}}=\frac 1f
\end{equation}
}and the magnification M$_1$ given by

\begin{equation}  \label{17}
M_1=\frac{S_1^{\prime }+(\frac{\omega _s}{\omega _i}-1)\:D_{CI_1}}S
\end{equation}
and{\it \ 
\begin{equation}  \label{18}
\frac 1{S+2D_{OM}}+\frac 1{S_2^{\prime }+(\frac{\omega _s}{\omega _i}-1)%
\:D_{CI_2}}=\frac 1f
\end{equation}
} the magnification M$_2$ given by

\begin{equation}  \label{19}
M_2=\frac{S_2^{\prime }+(\frac{\omega _s}{\omega _i}-1)\:D_{CI_2}}S
\end{equation}
where D$_{OM}$ is the distance from object to the classical mirror M (see
Fig. 3). The {\it proof of} {\it (L.4)} is similar to that of{\it \ (L.3)}
and here will be omitted.

\section{\bf Experimental tests for the geometric SPDC-quantum optics}

For an experimental test of {\it the Gauss like law of the thin lens
combined with a plane SPDC-QM }we propose an experiment based on a detailed
setup presented in Fig. 4 and in the optical geometric configuration shown
in Fig. 2b. Then, we predict that the image I of the object P (illuminated
by a high quality signal laser SL with s{\it ($\omega _s,{\bf k_s,e_s,\mu
_s))}$} will be observed in the idler beam, i{\it ($\omega _i,{\bf %
k_i,e_i,\mu _i)\equiv }$}i($\omega _p$-$\omega _s$,{\bf k$_p$-}{\bf k$_s$,e$%
_s^{*}$,-$\mu _s$}), when distances lens-object (S), lens-crystal-image
plane (S'), crystal-image plane (D$_{CI})$ and focal distance f of lens,
satisfy {\it thin lens+QM law (10). }Moreover{\it , }if {\it thin lens+QM
law (10)} is satisfied, the image I of that object P can be observed even
when instead of the signal source SL we put a detector D$_s.$ This last
statement is clearly confirmed recently, in the degenerate case $\omega
_s=\omega _i=\omega _p/2,$ by a remarkable {\it two-photon imaging experiment%
} [8]. Indeed, in these recent experiments, inspired by the papers of
Klyshko et al (see refs. quoted in [9]), was demonstrated some unusual {\it %
two-photon effects}, which looks very strange from classical point of view.
So, in these experiments, an argon ion laser is used to pump a nonlinear BBO
crystal ($\beta -BaB_2O_4)$ to produce pairs of {\it orthogonally polarized
photons} (see Fig. 1 in ref. [8] for detailed experimental setup). After the
separation of the {\it signal} and{\it \ idler }beams, an aperture (mask)
placed in front of one of the detectors (D$_s$) is illuminated by the {\it %
signal beam }through a convex lens. The surprising result consists from the
fact that an image of this aperture is observed in coincidence counting rate
by scanning the other detector (D$_i$) in the transverse plane of the idler
beam, even though both detectors single counting rates remain constants. For
understanding the physics involved in their experiment they presented an
''equivalent '' scheme ( in Fig. 3 in ref. [8]) of the experimental setup.
By comparison of their ''scheme'' with our optical configuration from Fig.
2b we can identify that the observed validity of the {\it two-photon} {\it %
Gaussian thin-lens equation}

\begin{equation}  \label{20}
\frac 1f=\frac 1S+\frac 1{S^{\prime }}
\end{equation}
as well as of the{\it \ linear magnification} 
\begin{equation}  \label{21}
M_0=\frac{S^{\prime }}S=2
\end{equation}
can be just explained by our results on the two-photon geometric law
(10)-(11) {\it of the thin lens combined with a plane SPDC-QM }for the
degenerate case $\omega _s=\omega _i=\omega _p/2.$ Therefore, the general
tests of the predictions (10)-(11) using a setup described in Fig. 4, are of
great importance not only in measurements in presence of the signal laser LS
(with and without coincidences between LS and idler detector D$_i),$ but
also in the measurements in which instead of the laser LS we put the a
signal detector D$_s$ in coincidence with D$_i.$ \ 

\section{\bf Conclusions}

In this paper the class of the {\it SPDC-phenomena} (1) is enriched by the
introducing the {\it crossing symmetric} {\it SPDC-processes }(2)-(3)
satisfying the same energy-momentum conservation law (4). Consequently, the
kinematical correlations (4)-(8) in conjunction with the Snellius relations
(7) allow us to introduce a new kind of optical devices called {\it quantum
mirrors. }Then, some laws of the {\it quantum mirrors, }such as:{\it \ }law
(9) of {\it thin plane SPDC-quantum mirror, }the {\it law} (10)-(11) {\it of
the thin lens combined with a plane SPDC-QM, }as well as,{\it \ the laws
(16)-(19),} {\it \ }are proved. These results are natural steps towards a 
{\it new geometric optics }which can be constructed for the kinematical
correlated SPDC-photons. In particular, the results obtained here are found
in a very good agreement with the recent results [8] on {\it two-photon
imaging experiment}. Moreover, we recall that {\it \ }all the results
obtained in the {\it two-photon ghost interference-diffraction} experiment
[6] was recently explained by using the concept of {\it quantum mirrors (}%
see Ref. [3]).

Finally, we note that all these results can be extended to the case of the 
{\it spherical quantum mirrors. }Such results, which are found in excellent
agreement to the recent experimental results [7] on {\it two-photon
geometric optics}, will be presented in a future publication. \ 
(This paper was published in Romanian Journal of Physics, Vol.45, P. 15, Bucharest 2000)

\bigskip \ \ \ \ 
\newpage
{\bf Figures}

\includegraphics[scale=1]{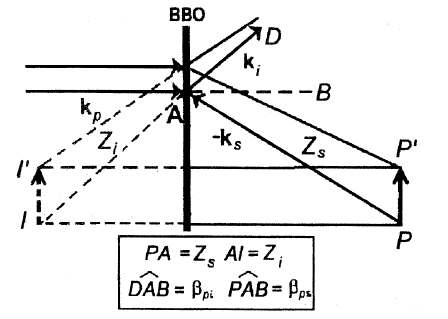} 
\begin{description}
\item[Fig. 1:]  The basic optical configuration of a{\it \ plane
SPDC-quantum mirror.}\ 
\end{description}
\includegraphics[scale=1]{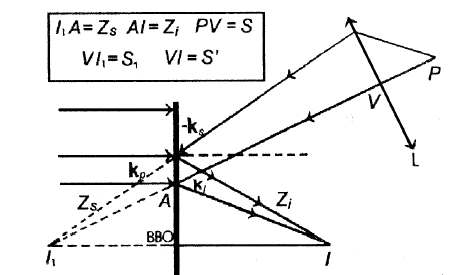}
\begin{description}
\item[Fig. 2a:]  The basic optical configuration for usual lens combined
with a {\it plane SPDC-quantum mirror.}\ 
\end{description}
\newpage
\includegraphics[scale=1]{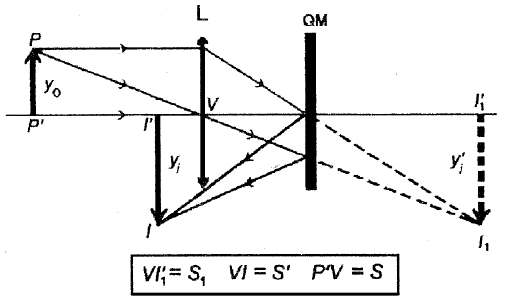}
\begin{description}
\item[Fig. 2b:]  The basic optical configuration for a proof of
magnification factor for a usual lens combined with a {\it plane} {\it %
SPDC-quantum mirror}.
\end{description}
\includegraphics[scale=1]{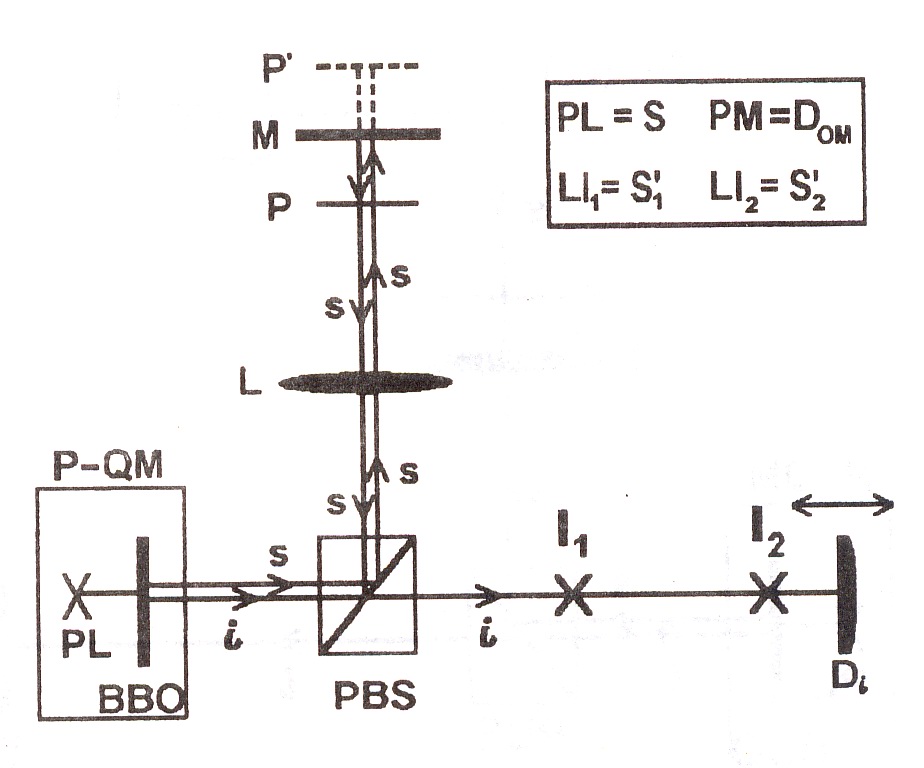}
\begin{description}
\item[Fig. 3:]  The basic optical configuration for usual lens combined with
a {\it plane} {\it SPDC-quantum mirror }and with{\it \ }a{\it \ classical
mirror}.
\end{description}
\includegraphics[scale=1]{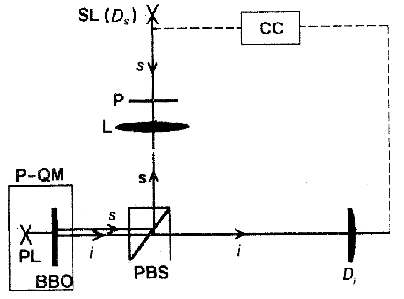}
\begin{description}
\item[Fig. 4:]  The scheme of the experimental setup for a test of the
geometric optics of correlated photons. The QM indicates the SPDC- {\it %
quantum mirror}, PBS is a polarization beam splitter, SL is a signal laser,
P is an object, L a convergent lens, D$_{i}$ is an idler detector and CC is
the coincidence circuit.
\end{description}

\end{document}